\documentclass{aa}
\usepackage{graphicx}
\usepackage{txfonts}
\usepackage{xcolor}

\begin{document} 

   \title{LCS: A Learnlet-Based Sparse Framework for Blind Source Separation}

   \author{V. Bonjean
          \inst{1}
          \and
          A. Gkogkou
\inst{1} \and J.-L. Starck
          \inst{1,2}
          \and
          P. Tsakalides
          \inst{1}
          }

   \institute{Institutes of Computer Science and Astrophysics, Foundation for Research and Technology Hellas (FORTH), Greece\\
              \email{victor.bonjean40@gmail.com}
         \and
             Université Paris-Saclay, Université Paris Cité, CEA, CNRS, AIM, 91191, Gif-sur-Yvette, France}

   \date{Received XXX; accepted XXX}
\abstract{
Blind source separation (BSS) plays a pivotal role in modern astrophysics by enabling the extraction of scientifically meaningful signals from multi-frequency observations. Traditional BSS methods, such as those relying on fixed wavelet dictionaries, enforce sparsity during component separation, but may fall short when faced with the inherent complexity of real astrophysical signals. In this work, we introduce the Learnlet Component Separator (LCS), a novel BSS framework that bridges classical sparsity-based techniques with modern deep learning. LCS utilizes the Learnlet transform—a structured convolutional neural network designed to serve as a learned, wavelet-like multiscale representation. This hybrid design preserves the interpretability and sparsity-promoting properties of wavelets while gaining the adaptability and expressiveness of learned models. The LCS algorithm integrates this learned sparse representation into an iterative source separation process, enabling effective decomposition of multi-channel observations. While conceptually inspired by sparse BSS methods, LCS introduces a learned representation layer that significantly departs from classical fixed-basis assumptions. We evaluate LCS on both synthetic and real datasets, demonstrating superior separation performance compared to state-of-the-art methods (average gain of about 5 dB on toy model examples). Our results highlight the potential of hybrid approaches that combine signal processing priors with deep learning to address the challenges of next-generation cosmological experiments.
}

\maketitle

\section{Introduction}

The separation of astrophysical components from multi-frequency sky observations is a fundamental task in observational cosmology and astrophysics. It enables the isolation of a source signal from a complex mixture of astrophysical foregrounds and instrumental noise. A prominent example is the extraction of the Cosmic Microwave Background (CMB) from multi-band observations, which played a central role in the success of the Planck mission \citep{planck}. Traditional component separation methods in this context include Principal Component Analysis (PCA), which removes the most dominant modes in the frequency domain \citep{pca}, and ICA-based approaches \citep{ica}, which exploit the statistical independence of components. While simple and model-agnostic, these methods risk partial loss of the cosmological signal. More sophisticated techniques were developed for Planck including internal linear combination techniques like NILC \citep[Needlet Independent Linear Combination,][]{nilc} and MILCA \citep[Modified Internal Linear Combination Algorithm,][]{milca}, blind source separation methods such as SMICA \citep[Spectral Matching ICA,][]{smica}, self-supervised deep learning methods \citep{bonjean2024}, and sparse techniques like GMCA \citep[Generalised Morphological Component Analysis,][]{gmca}. These methods disentangle sources from multi-channel observations by leveraging assumptions such as statistical independence, spectral diversity, or sparse representability.

In particular, sparse component separation has proven to be a powerful approach for modeling structured, non-Gaussian emissions such as foregrounds exhibiting filamentary or localized morphologies. GMCA exploits the assumption that astrophysical components can be sparsely represented in a certain transform domain. The typically used domains are wavelets and specifically starlet, biorthogonal, or mexican hat and they are all chosen based on the complexity of the image under study. By promoting sparsity via morphological diversity, GMCA has been shown to achieve accurate CMB and foreground reconstruction in both simulated and real data in astrophysical applications and beyond \citep{gmca_cmb, gmca_sz, gmca_eeg, gmca_remsens}. However, a key limitation stems from the fixed nature of the wavelet dictionary: while efficient and interpretable, it may not optimally capture the statistical structure of all signal types or adapt to data variability, particularly in non-Gaussian or non-stationary regimes.

In astrophysics, the need for efficient and robust component separation techniques extends beyond the CMB. With the upcoming deployment of radio interferometers like the Square Kilometre Array \citep[SKA,][]{ska}, there is a renewed interest in separating diffuse emission from target cosmological signals. In particular, HI intensity mapping aims to detect the 21-cm emission from neutral hydrogen across cosmic time, offering a promising probe of large-scale structure, dark energy, and the epoch of reionization \citep{ska_cosmo, ska_cosm}. However, this faint cosmological signal is contaminated by foregrounds several orders of magnitude brighter, including Galactic synchrotron and free-free emission, as well as extragalactic point sources \citep{alonso2015}. Effective foreground cleaning is thus essential for recovering the HI signal with high fidelity. Recent works \citep[][Gkogkou et al., in prep.]{carucci2020, makinen2021, matshawule2021, marta2022, carucci2024, mertens2024} have explored deep learning-based priors, regularization schemes, and learned sparse representations to tackle the challenge of foreground removal in simulated SKA data.

In parallel, the field of signal processing and machine learning has made a significant progress in data-driven sparse modeling. A classical approach in sparse coding is the Iterative Shrinkage-Thresholding Algorithm \citep[ISTA,][]{fista}, which solves linear inverse problems with sparsity constraints by iteratively applying a soft-thresholding operation to promote sparse solutions. While ISTA is based on convex optimization and offers well-characterized convergence guarantees, its iterative nature often results in slow convergence in practice. To address this, \cite{lista} introduced the Learned ISTA (LISTA), a neural network architecture that unrolls the ISTA iterations into a fixed-depth feedforward network. By learning the optimal parameters of each iteration (such as the dictionary and thresholding functions) from data, LISTA achieves significantly faster convergence and better adaptability to specific signal classes, while maintaining the interpretability of sparse coding. Building upon these ideas, \cite{lpalm} developed LPALM (Learned Proximal Alternating Linearized Minimization), a hybrid framework that combines deep supervised learning with optimization-based sparse component separation. LPALM extends the principles of LISTA by integrating it into a proximal optimization loop tailored for separating astrophysical sources, demonstrating promising results in the identification of elemental components in supernova remnants on X-Ray data.

Expanding on the concept of hybrid methods offering both interpretability and learning capability, we investigate the use of neural networks designed to enhance classical signal transforms. Among them, the Learnlet transform \citep{learnlet} introduces a neural network architecture designed to emulate wavelet-like behavior while learning filters and thresholding parameters directly from data. This approach preserves the interpretability and multiscale structure of classical wavelets, while offering greater adaptability to complex signal distributions. Unlike traditional wavelet-based dictionaries, the learned filters - the learnlets - can adapt to complex and noisy environments, improving the robustness of source separation. While they have already demonstrated effectiveness in deconvolving astronomical images \cite{utsav}, their integration into structured component separation frameworks remains largely unexplored.

In this work, we introduce the Learnlet Component Separator (LCS), a novel BSS algorithm that embeds a learned sparse representation—called the Learnlet transform—into an iterative separation procedure. The Learnlet transform is a structured convolutional neural network (CNN) designed to emulate a wavelet-like multiscale behavior while allowing for data-driven adaptation. Its architecture preserves key characteristics of traditional wavelet bases, such as locality, multiresolution decomposition, and interpretability, while improving upon them through learned filter banks optimized via deep learning for the specific signal class.

The LCS algorithm iteratively estimates the sources that are mixed in each frequency channel by leveraging the sparsity-promoting capabilities of the Learnlet transform within a sources-mixing matrix estimation loop. This design allows LCS to combine the robustness and theoretical grounding of sparse representation methods with the adaptability and expressive power of deep neural networks. In this paper, we describe the architecture of LCS, examine its relationship to classical methods like GMCA, and assess its performance through extensive numerical experiments including toy models with structured scenes (airplane, boat, Barbara)  and real-world astrophysical data (X-ray supernova remnants and the extraction of CMB and Sunyaev-Zel'dovich—SZ—signals).

Our findings demonstrate that LCS consistently outperforms conventional BSS techniques, especially in complex scenarios involving nonstationary foregrounds or non-ideal mixing conditions (such as in the presence of noise). These results highlight the potential of LCS for future applications to SKA data, where accurate foreground removal will be essential for an accurate extraction of the HI signal in cosmology.

The remainder of this paper is structured as follows. In Sect.~\ref{sec:sec1}, we present in detail the Learnlet transform, its architecture and its training on the ImageNet dataset. Section \ref{sec:sec2} describes the main LCS algorithm for source separation, which incorporates the Learnlet transform. Section \ref{sec:sec3} provides experimental results and comparative evaluations across four distinct experiments involving both toy models and astrophysical images. Finally, Section~\ref{sec:sec4} concludes the paper with a discussion of the methods, their relevance to astrophysical applications, and potential future directions.

\begin{figure*}[!ht]
    \centering
    \includegraphics[width=0.9\linewidth]{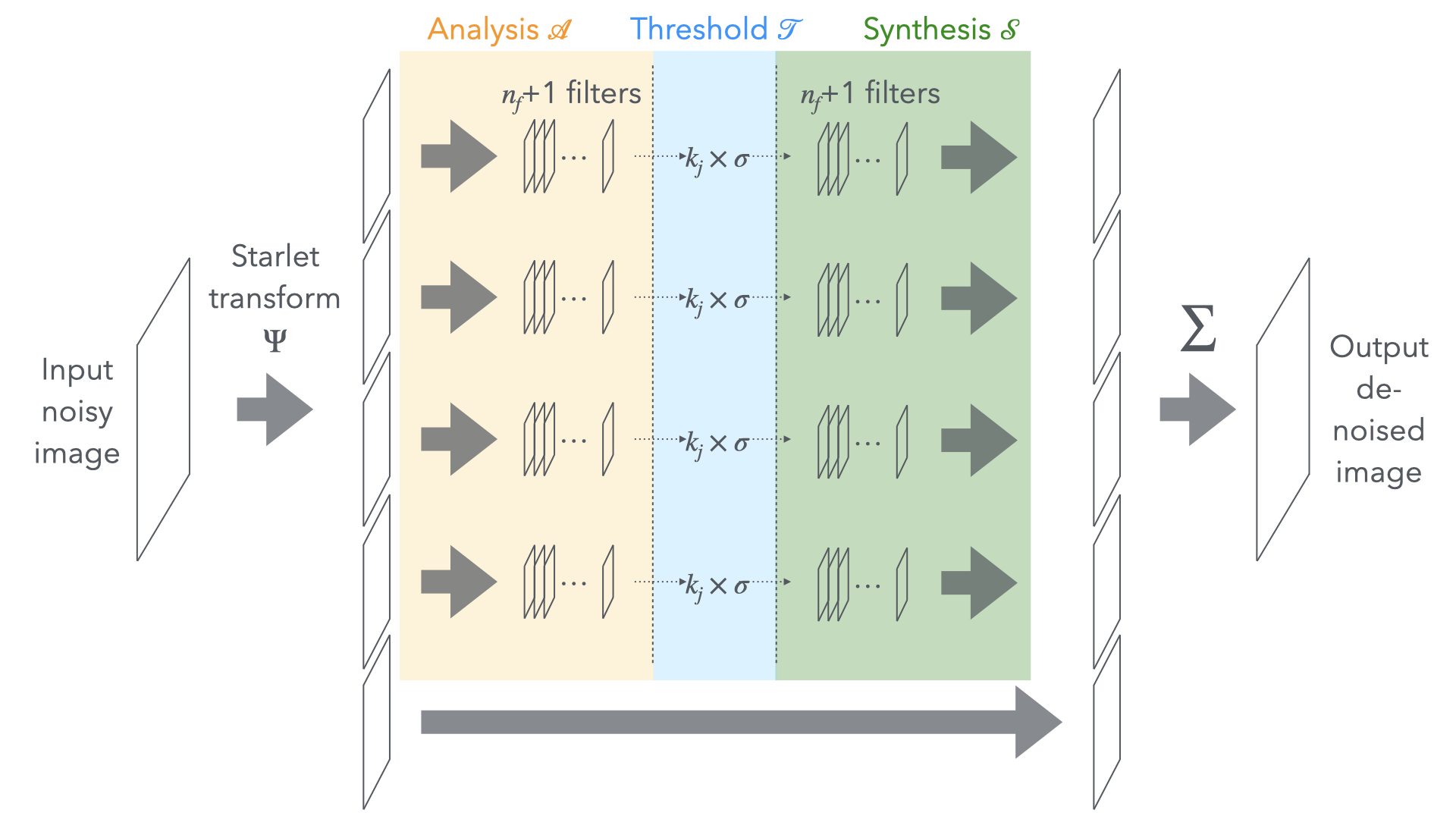}
    \caption{Architecture of the Learnlet network: Example with a number of scales $J=5$. The $n_f$ first filters of the $J-1$ first scales of both the analysis $\mathcal{A}$ and of the synthesis $\mathcal{S}$ layers are learned, as well as the $J-1$ thresholding parameters $k_j$ of the threshold layer $\mathcal{T}$.}\label{fig:fig1}
\end{figure*}

\section{Learnlets}\label{sec:sec1}

\subsection{Architecture}

Learnlet is a transform designed for image denoising, built upon a specialized Convolutional Neural Network (CNN) architecture, inspired by wavelet transforms and leveraging the representational power of deep learning, as originally introduced in \cite{learnlet}. Compared to other state-of-the-art deep learning denoisers like UNets \citep{unet}, Learnlet offers greater interpretability. Its learned filters—referred to as learnlets—form a sparse basis, enabling the imposition of sparsity in subsequent tasks. The transform takes as input a noisy image $\mathbf{X}=\mathbf{Y}+\mathbf{N}$, where $\mathbf{Y}$ is the ground truth image and $\mathbf{N}$ is a realization of noise with standard deviation $\sigma$, and estimates a denoised image $\hat{\mathbf{Y}}$. The architecture is structured as follows:

\begin{itemize}
    \item We begin by applying the starlet transform \citep{starlet} to the input image $\mathbf{X}$, $\alpha = \Psi \mathbf{X}$, decomposing it into $J$ distinct scales with coefficients $\alpha_j$. In our case, $J=5$.
    \item The first $J-1$ scales are sequentially passed through an analysis layer $\mathcal{A}$, a thresholding layer $\mathcal{T}$, and a synthesis layer $\mathcal{S}$. The analysis layer $\mathcal{A}$ consists of $n_f+1$ filters with kernel size $h \times w$, producing convolved outputs with the same variance as the input. To ensure this variance consistency, the filters are normalized by the sum of their squared coefficients. A total of $n_f \times (J-1)$ filters are learned, while the remaining $J-1$ filters (indexed as $n_f+1$) are fixed to the identity filter $\delta$. In our implementation, we set $h=w=5$ and $n_f=64$.
    \item Next, a thresholding operation $\mathcal{T}$ is applied to the output feature maps from the analysis layer for each of the $J-1$ scales $\mathcal{A}$($\alpha_j$). We employ hard thresholding, using a threshold of $k_j \times \sigma$ at each scale $j$, where $\sigma$ denotes the standard deviation of the input noise. The $J-1$ threshold parameters $k_j$ for each scale are also learned during training and are constrained within the range 0 < $k_j$ < 5.
    \item The thresholded feature maps $\mathcal{T}$($\mathcal{A}$($\alpha_j$), $k_j \times \sigma$) are then passed to the synthesis layer $\mathcal{S}$, which also consists of $n_f+1$ filters of size $h\times w$ for each of the first $J-1$ scales. Of these, $n_f\times (J-1)$ filters are learned, while the remaining filters $n_f+1$ are set to $\delta - \sum_{i=1}^n$ ${{\mathcal{F}_\mathcal{S}}_j}_i$*${{\mathcal{F}_\mathcal{A}}_j}_i$, where ${{\mathcal{F}_\mathcal{S}}_j}_i$ and ${{\mathcal{F}_\mathcal{A}}_j}_i$ 
    represent the $i$-th synthesis and analysis filters at scale $j$, respectively. This construction ensures that the sum of the outputs from $\mathcal{S}$($\mathcal{T}$($\mathcal{A}$($\alpha_j$), $k_j \times\sigma$)) exactly reconstructs the input when $\sigma=0$, guaranteeing perfect reconstruction in the absence of noise.
    \item The final denoised map $\hat{\mathbf{Y}}$ is obtained by summing the denoised outputs from the first $J-1$ scales along with the unchanged coarse scale $\alpha_J$. This is expressed as:
    \begin{equation}
        \hat{\mathbf{Y}}=\alpha_J + \sum_{j=1}^{J-1} \mathcal{S}(\mathcal{T}(\mathcal{A}(\alpha_j), k_j \times\sigma)).
    \end{equation}
\end{itemize}

A schematic representation of the Learnlet architecture is shown in Fig.~\ref{fig:fig1}. We reimplemented the original Learnlet algorithm from \cite{learnlet} in PyTorch to enable more efficient GPU utilization. Additionally, we created a user-friendly and well-documented Python class, along with example notebooks, all available on GitHub\footnote{\url{https://github.com/vicbonj/learnlet.git}}.

\begin{figure}[!ht]
    \centering
    \includegraphics[width=1\linewidth]{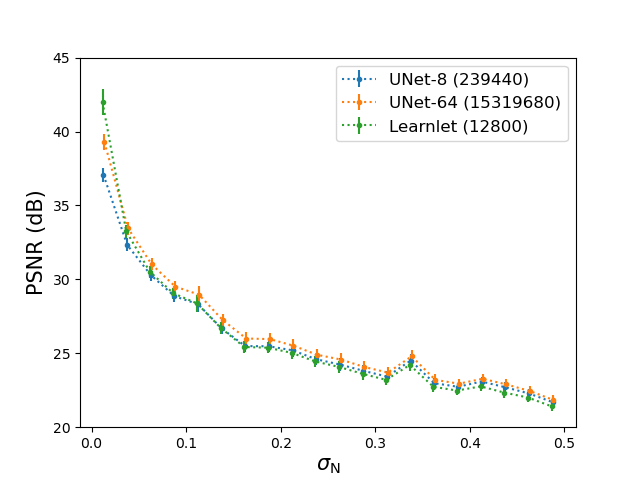}
    \caption{PSNR of the denoised test images for Learnlet (green), UNet-8 (blue), and UNet-64 (orange) plotted against the input noise level 
    $\sigma_\mathrm{N}$. The legend includes the number of trainable parameters for each network.}
    \label{fig:fig2}
\end{figure}

\subsection{Training on ImageNet}\label{sec:imagenet}

The learnlets must first be trained. To achieve this, we selected a subset of the ImageNet database \citep{imagenet}, which offers a diverse range of images with rich and complex features—ideal for training generalizable and robust learnlets. The dataset is publicly accessible via their website\footnote{\url{https://www.image-net.org}}. Specifically, we used the first 10.000 images from the ImageNet test set, converting them from RGB to gray scale (with pixel values normalized between 0 and 1) and resizing them to $256 \times 256$ pixels. The dataset was then split into 8.000 images for training, 1.000 for validation, and 1.000 for testing.

For all sets, we added synthetic Gaussian noise $\mathbf{N}$ to the images, with the noise standard deviation $\sigma_\mathrm{N}$ randomly sampled between 0 and 0.5. This resulted in a collection of noisy images $\mathbf{X}$ and their corresponding clean counterparts $\mathbf{Y}$, such that where $\mathbf{X} = \mathbf{Y} + \mathbf{N}$. 

We trained the Learnlet model using $\mathbf{X}$ as input and $\mathbf{Y}$ as the corresponding ground truth output. The network was configured with $n_f=64$ filters of size $5\times5$, differing from the original setup in \cite{learnlet}, but yielding improved and faster performance. We removed all bias parameters from the convolutional layers. We used an Adam optimizer with a learning rate $\ell_\mathrm{r}=10^{-4}$ and mean squared error (MSE) as the loss function. The training ran on an NVIDIA A100 GPU and completed in approximately 8 hours.

The performance of the Learnlet model on the test set is shown in green in Fig.~\ref{fig:fig2}. The plot displays the average Peak Signal-to-Noise Ratio (PSNR, defined in the next section) values (in dB) of the reconstructed images as a function of the input noise standard deviation $\sigma_\mathrm{N}$. For comparison, we evaluate two UNet architectures \citep{unet}: a lightweight version (UNet-8, shown in blue) initialized with 8 filters, and a more complex variant (UNet-64, shown in orange) initialized with 64 filters.  The numbers in parentheses in the legend indicate the total number of trainable parameters in each model, highlighting the markedly lower complexity of Learnlet (12,800 parameters) compared to the UNet architectures, which range from 239,440 (UNet-8) to 15,319,680 (UNet-64).

Overall, the Learnlet model delivers strong performance, matching that of a small UNet-8 and even outperforming the larger UNet-64 at very low values of $\sigma_\mathrm{N}$, thanks to its ability to achieve exact reconstruction. Although learnlets may perform slightly worse than a UNet, they offer several advantages: they are fully trustworthy, provide exact reconstruction, are mathematically well defined as frames, inherit the convergence properties of wavelets in proximal iterative algorithms, and require $\sim100$ times fewer parameters. Hence, Learnlets provide an interesting trade-off between efficiency and robustness. 

In the appendix, we show an example of learned filters and the evolution of the thresholding parameters as a function of $\sigma_\mathrm{N}$.

The learnlets form an optimized sparse basis tailored to natural image structures. Building on this foundation, we introduce a Blind Source Separation algorithm called LCS, which shares structural similarities with GMCA but incorporates key novelties, which are discussed in the following sections.

\section{The Learnlet Component Separator (LCS)}\label{sec:sec2}

In this section, we describe in detail the algorithm LCS in the context of BSS.

\begin{figure*}[!ht]
    \centering
    \includegraphics[width=0.49\linewidth]{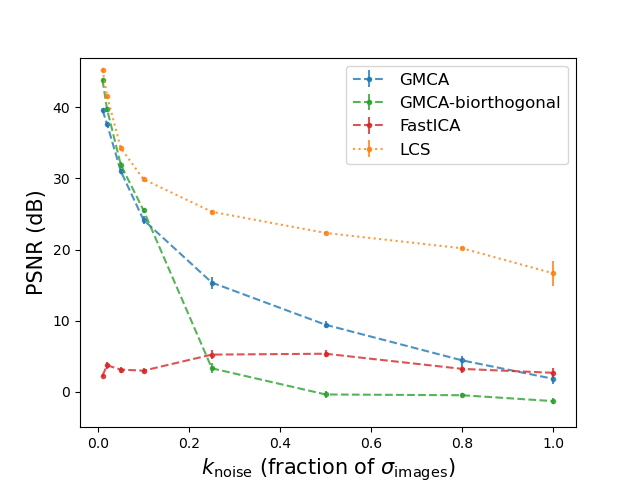}
    \includegraphics[width=0.49\linewidth]{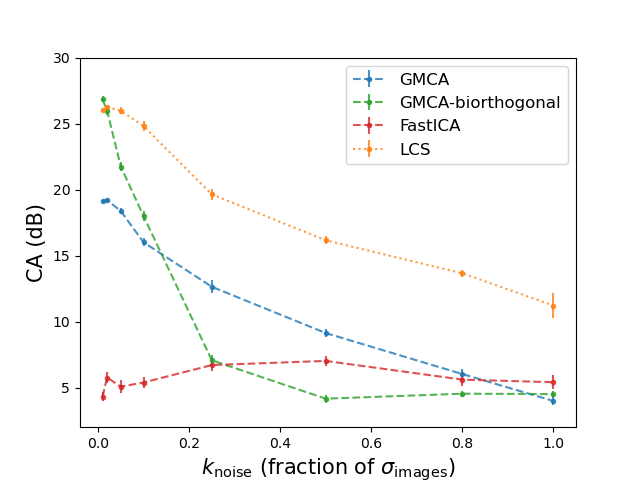}
    \caption{Performance comparison of various BSS methods on standard test images: LCS (orange), GMCA (blue), GMCA-biorthogonal (green), and FastICA (red). Left panel: mean PSNR values on the source matrix $\mathbf{S}$ as a function of the input noise level $\mathrm{k}_\mathrm{noise}$. Right panel: corresponding mean Component Angle (CA) values on the mixing matrix $\mathbf{A}$. Each point represents the average and standard error over 10 independent realizations of both the mixing matrices A and the noise.} 
    \label{fig:fig3}
\end{figure*}

\subsection{Formalism of BSS}

Blind Source Separation is a core challenge in signal processing, aiming to recover unknown source signals $\mathrm{S}$ from observed mixtures $\mathbf{Y}$ that may also include noise $\mathbf{N}$, all without prior knowledge of the mixing process $\mathbf{A}$. Mathematically, this can be expressed as:

\begin{equation}
    \mathbf{Y} = \mathbf{A} \cdot \mathbf{S}+\mathbf{N}.
\end{equation}

BSS has broad applications across multiple fields, including biomedical signal analysis (e.g., EEG and fMRI), audio processing (e.g. speech separation), and notably in astrophysical imaging, for tasks such as CMB and SZ extraction, HI signal recovery, and source detection.

\subsection{LCS main algorithm}

The algorithm iteratively estimates both the sources, denoised using the Learnlet transform, and the mixing matrix, drawing inspiration from GMCA. LCS employs an iterative scheme that alternates between updating the source estimates and refining the mixing matrix:

\begin{itemize}
    \item \textbf{$\mathbf{S}$ estimate}: We first estimate $\hat{\mathbf{S}}$ with a fixed estimate of the mixing matrix $\hat{\mathbf{A}}$, by minimizing the equation:
    
    \begin{equation}
         \min_{{\mathbf{S}}} ||\mathbf{Y} - \hat{\mathbf{A}}  \cdot \mathbf{S}||_2^2 + \lambda ||{\mathcal{L}_{\mathbf{S}}}||_\mathrm{p},
    \end{equation}
    
    where the second term is obtained, with p=0, by performing a $\ell_0$-norm constraint on $\mathbf{S}$ to promote sparsity. This is done by the hard-thresholding operation from the Learnlet transform $\mathcal{L}_{\mathbf{S}_i}$ for each source component $\mathbf{S}_i$.
     \item \textbf{$\mathbf{A}$ estimate}: the estimate of the mixing matrix, $\hat{\mathbf{A}}$, is refined by minimizing the equation:

     \begin{equation}
         \min_{{\mathbf{A}}} ||\mathbf{Y}^{\mathrm{no-coarse}} - \mathbf{A}  \cdot {\mathcal{L}_{\hat{\mathbf{S}}}}^{\mathrm{no-coarse}}||_2^2,
    \end{equation}
     
     where the coarse scales have been removed to eliminate low-frequency spatially correlated components that could otherwise introduce ambiguity and degrade the quality of source separation.
\end{itemize}

Convergence is typically reached when successive iterations produce only minimal changes in all $\hat{\mathbf{S}}$, indicating that both the source separation and the estimate of $\mathbf{A}$ have stabilized.

After the minimization has converged, an additional post-processing step is performed to denoise each estimated source $\hat{\mathbf{S}}_i$ using the Learnlet transform as a denoiser, with the noise level estimated individually for each component $i$. The noise level is derived from the known noise covariance $\Sigma_{\mathbf{N}}$ and the pseudo-inverse of the estimated mixing matrix $\hat{\mathbf{A}}$, leading to the computation of the source covariance: 

\begin{equation}
    \Sigma_{\hat{\mathbf{S}}} = {\hat{\mathbf{A}}}^{+} \Sigma_{\mathbf{N}} ({\hat{\mathbf{A}}}^{+})^\mathsf{T}.
\end{equation}

The main differences of LCS and GMCA are as follows:

\begin{itemize}
    \item \textbf{Fixed Thresholding:} In LCS, the threshold is fixed, eliminating the need for a threshold decay strategy, which in GMCA often depends on the image type. We tried several multiplicative factor of the $\sigma_\mathrm{NMAD}$ of the sources and found that $1.5 \times \sigma_\mathrm{NMAD}$ gives the best results for most of our applications.
    \item \textbf{Hard vs. Soft Thresholding:} LCS employs hard-thresholding instead of the soft-thresholding used in GMCA, thereby avoiding bias introduced by the latter.
    \item \textbf{Source Update Domain:} LCS updates the source estimates $\hat{\mathbf{S}}$ in the real space, whereas GMCA performs this update in the wavelet domain—an approximation if the decomposition is not orthogonal.
    \item \textbf{Post-processing Denoising Step:} LCS includes an additional final denoising step applied to the estimated sources $\hat{\mathbf{S}}$.
\end{itemize}

The learnlets ${\mathcal{L}_{\hat{\mathbf{S}}}}_i$ are individually trained for each of the component $i$ to enhance separation quality by learning optimal bases and thresholding parameters tailored to the morphology, features, and contrasts of each component. However, a shared basis $\mathcal{L}$ can also be used for all components. In this work, we explore the use of such a common basis by employing Learnlets pre-trained on ImageNet (see Sect.~\ref{sec:imagenet}) as a general-purpose option. Additionally, we investigate component-specific training using numerical simulations or texture datasets, leveraging transfer learning with weights initialized from the ImageNet-trained models.

\subsection{Quality assessment}

For all experiments presented in this paper, we assess the quality of the source reconstruction $\mathbf{S}$ using the Peak Signal-to-Noise Ratio (PSNR) in decibels (dB). PSNR is a widely used metric in signal processing that quantifies the accuracy of image reconstruction and is defined as:

\begin{equation}
    \mathrm{PSNR} = 10 \cdot \log_{10} \left( \frac{\mathrm{max^2}}{\mathrm{MSE}}\right),
\end{equation} 
where max denotes the maximum possible pixel value in the image (set to 1 in our case), and MSE represents the mean squared error between the estimated sources $\hat{\mathbf{S}}$ and the ground truth sources $\mathbf{S}$.

The quality of the mixing matrix reconstruction is evaluated using the Criterion on $\mathbf{A}$ (CA), which measures how closely the product of the pseudoinverse of the estimated mixing matrix $\hat{\mathbf{A}}$ with the true matrix $\mathbf{A}$ approximates the identity matrix. Ideally, this product satisfies $\hat{\mathbf{A}}^+ \cdot \mathbf{A} = \mathbf{I}$, where $\mathbf{I}$ is the identity matrix. The CA metric, expressed in decibels (dB), is defined as:
\begin{equation}
    \mathrm{CA} = -10 \cdot \log_{10} \left( \frac{1}{\mathrm{N}} \left\| \hat{\mathbf{A}}^{+} \cdot \mathbf{A} - \mathbf{I} \right\|_F \right),
\end{equation}
where N is the number of channels and $||\cdot ||_F$ denotes the Frobenius norm.

\section{Numerical experiments}\label{sec:sec3}

In this paper, we conduct four distinct experiments to evaluate the source separation capabilities of Learnlets. Two experiments use a shared, general basis $\mathcal{L}$ derived from Learnlets trained on ImageNet (see Sect.~\ref{sec:imagenet}), while the other two use individually retrained bases $\mathcal{L}_i$ tailored to specific components. We evaluate both approaches, on standard test photographic images and on astrophysical imagery.

\subsection{Experiments using a common basis $\mathcal{L}$}

\subsubsection{Experiments with standard test images}
\label{sec:toy}

In this experiment, the target source signal $\mathbf{S}$ consists of three components, represented by grayscale images of Barbara, an airplane, and a boat, common reference points in signal processing for denoising tasks. Each image has a resolution of $256 \times 256$ pixels and is normalized to a [0, 1] range.
We perform multiple component separation runs using a randomly generated mixing matrix $\mathbf{A}$, sampled from a uniform distribution in [0, 1] with 6 channels, and constrained to have a condition number below 5. The noise matrix $\mathbf{N}$ varies between realizations. Specifically, we repeat the experiment 10 times for each noise level, where the noise in the observations $\mathbf{Y}$ is scaled as $k_\mathrm{noise} \times \sigma_\mathrm{images}$, with $k_\mathrm{noise}$ taking values in {0.01, 0.02, 0.05, 0.1, 0.25, 0.5, 0.8, 1}, and $\sigma_\mathrm{images}$ being the standard deviation of the original sources $\mathbf{S}$.
For each realization, we estimate $\hat{\mathbf{S}}$ and $\hat{\mathbf{A}}$ using LCS, and compare the results against different variants of GMCA: i) standard GMCA using starlets, and ii) GMCA-biorthogonal, which uses bi-orthogonal wavelets. We also include comparisons with the FastICA algorithm.

The results are presented in Fig.~\ref{fig:fig3}. The left panel displays the mean PSNR values for the source estimates $\mathbf{S}$ along with error bars, plotted against the input noise level $k_\mathrm{noise}$. The right panel shows the mean CA values for the mixing matrix estimates $\mathbf{A}$. LCS results are shown in orange and compared against GMCA (blue), GMCA-biorthogonal (green), and FastICA (red).

In the source reconstruction results shown on the left panel, LCS outperforms all GMCA variants by 3 to 5 dB in the low-noise regime ($k_\mathrm{noise} < 0.1$), while FastICA performs poorly given the specific image and channel configuration. As the noise level increases ($k_\mathrm{noise} > 0.1$), LCS delivers a substantial improvement, achieving gains of 10 to 15 dB over the other methods.

In the mixing matrix reconstruction results shown in the right panel, LCS surpasses all other methods by 5 to 15 dB across most noise levels. The only exception is at the lowest noise level ($k_\mathrm{noise} = 0.01$), where GMCA-biorthogonal also performs exceptionally well, slightly outperforming LCS by about 1 dB.

\subsubsection{Experiments with astrophysical images}

\begin{figure*}[!ht]
    \centering
    \includegraphics[width=0.49\linewidth]{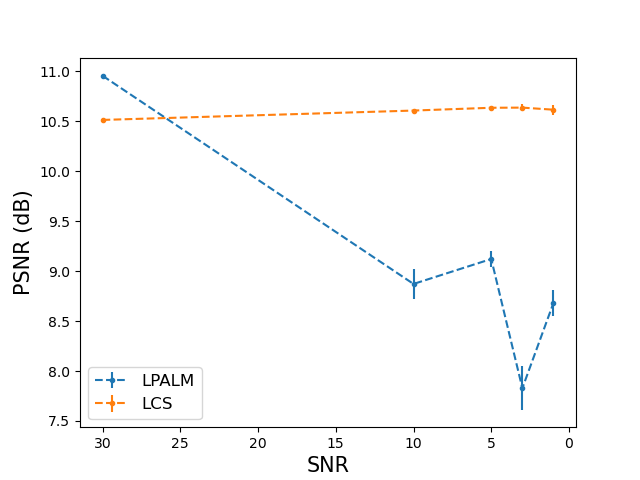}
    \includegraphics[width=0.49\linewidth]{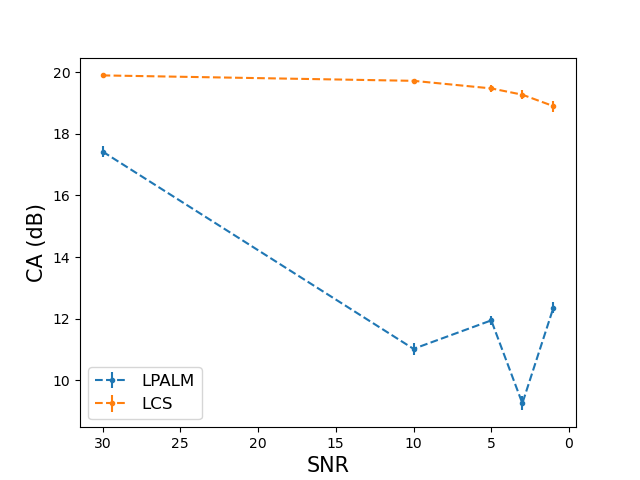}
    \caption{Performance comparison of different BSS methods in the astrophysical case, evaluated as a function of input noise level expressed in terms of SNR: LCS is shown in orange, and LPALM in blue. Each data point represents the mean and standard error over 150 realizations from the test set of \cite{lpalm}. Left panel: average PSNR for the source estimates $\mathbf{S}$. Right panel: average CA for the mixing matrix estimates $\mathbf{A}$.}
    \label{fig:fig6}
\end{figure*}

\begin{figure*}[!ht]
    \centering
    \includegraphics[width=0.49\linewidth]{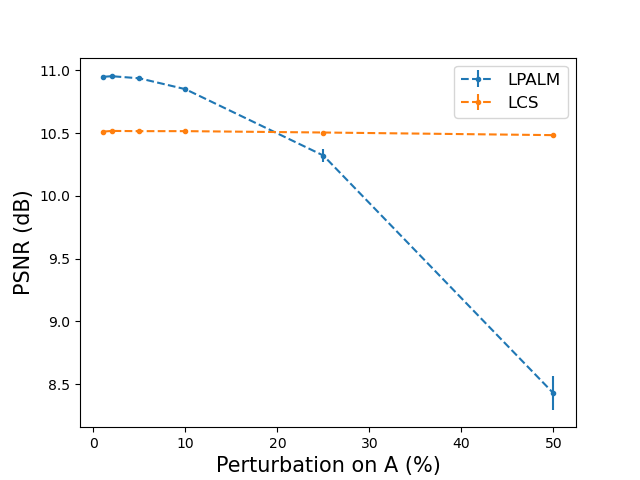}
    \includegraphics[width=0.49\linewidth]{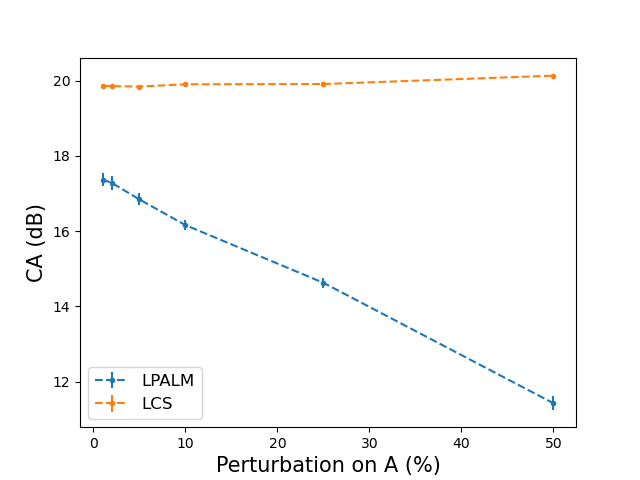}
    \caption{Comparison of BSS method performance in the astrophysical case at a fixed SNR of 30, as a function of the percentage of perturbation applied to the mixing matrix $\mathbf{A}$. LCS is shown in orange and LPALM in blue. Each point represents the mean and standard error across 150 realizations from the test set of \cite{lpalm}. Left panel: average PSNR for the source estimates $\mathbf{S}$. Right panel: average CA for the mixing matrix estimates $\mathbf{A}$.}
    \label{fig:fig7}
\end{figure*}

In this experiment, we use X-ray astronomical images from \cite{lpalm}, which are publicly available on their GitHub repository\footnote{\url{https://github.com/mfahes/LPALM.git}}. We include a comparison with their LPALM algorithm—a semi-supervised deep learning method for blind source separation that achieved state-of-the-art performance on this dataset. LPALM is a supervised approach in which the proximal operator is learned from paired mixtures and corresponding ground-truth sources using an unrolled, end-to-end optimization framework. The model is trained specifically to recover both the source components and the mixing matrix from mixed observations, assuming the mixing matrix is known during training. The dataset consists of four source components in $\mathbf{S}$ and 64 channels in the mixing matrix $\mathbf{A}$. We train LPALM across various noise levels, with SNR values ranging from 30 down to 1, and evaluate both LCS and LPALM across all noise realizations and mixing matrices from their test set, which includes 150 instances.

The results are presented in Fig.~\ref{fig:fig6}, with LCS shown in orange and LPALM in blue for comparison. PSNR and CA are plotted against the noise level, expressed as SNR. At a high SNR of 30, corresponding to very low noise, LPALM achieves slightly better performance, with about 0.5 dB higher PSNR—likely due to its image filtering strategy in the final iteration. However, as SNR decreases (i.e., noise increases), LPALM’s performance degrades noticeably, while LCS remains largely unaffected by the noise level in this setting, likely thanks to the high number of channels (64) available for source separation.

LCS consistently provides more accurate estimates of the mixing matrix $\mathbf{A}$, which is particularly notable given that LCS operates in a fully blind manner—unlike LPALM, which is semi-blind and explicitly trained for this type of data. In this experiment, the Learnlets used by LCS were not retrained and remained those originally computed from ImageNet. This highlights a key advantage of LCS: it performs well even with little or no task-specific training data. When training data are available for component-specific training with individual bases $\mathcal{L}_i$, the results can be further enhanced, as demonstrated in the following experiments.

In this experiment, we also introduce a small perturbation to the mixing matrix $\mathbf{A}$ during testing at SNR = 30 (corresponding to a low noise level in $\mathbf{Y}$) to illustrate the limitations of LPALM as a semi-supervised method and compare its performance with LCS. Since LPALM is trained to recover mixing matrices similar to those in its training set, we perturb $\mathbf{A}$ as follows: $\widetilde{\mathbf{A}}=\mathbf{A} + k\times\mathcal{N}(0, 1)\times\mathbf{A}$, where $k$ varies from 0.1 to 0.5 and $\mathcal{N}(0, 1)$ is a normal distribution with zero mean and unit variance. The resulting PSNR and CA values are shown in Fig.~\ref{fig:fig7}. As expected, LPALM’s performance begins to degrade significantly once the perturbation of $\mathbf{A}$ reaches about 20\%, impacting both PSNR and CA. In contrast, LCS—being a fully blind method that does not depend on supervision—remains unaffected by these perturbations. This test further underscores the robustness of LCS and highlights the advantage of a fully unsupervised approach: when the mixing matrix deviates even slightly from the training data, an unsupervised algorithm like LCS maintains stable performance without degradation.

\subsection{Experiments using individually retrained bases $\mathcal{L}_i$}

In the following experiments, instead of using Learnlets trained on ImageNet, we perform individual training for each source type in $\mathbf{S}$ to better optimize the induced sparsity according to the distinct morphologies and features of each component.

\subsubsection{Experiments with standard test images}

In this experiment, we train a separate Learnlet network for each source to be estimated, resulting in individual bases $\mathcal{L}_i$. For this toy model, we use the Describable Texture Dataset (DTD) from \cite{textures}, which is publicly available on their website\footnote{\url{https://www.robots.ox.ac.uk/~vgg/data/dtd/index.html}}. The dataset includes 47 texture categories (such as banded, dotted, spiralled, etc.), each containing 120 images. Here, we focus specifically on the banded and dotted texture classes.

Both networks, $\mathcal{L}_\mathrm{banded}$ and $\mathcal{L}_\mathrm{dotted}$, are initialized with the ImageNet-pretrained weights from the previous section and then fine-tuned independently. For each texture class, we use the first 119 images, presenting each image ten times with different noise realizations each time.

We define the target sources $\mathbf{S}$ as the two remaining (120th) images from the banded and dotted classes, which were not used during training. The subsequent steps follow the same procedure as in our experiments with the standard test images: for each noise level $\mathrm{k}_\mathrm{noise}$, we generate 10 realizations of both the mixing matrix $\mathbf{A}$ and the noise $\mathbf{N}$, and then benchmark LCS against GMCA, GMCA-biorthogonal, and FastICA.

\begin{figure*}[!ht]
    \centering
    \includegraphics[width=0.49\linewidth]{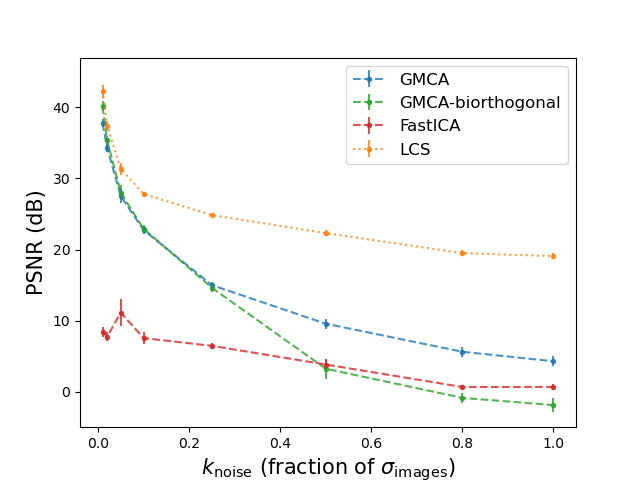}
    \includegraphics[width=0.49\linewidth]{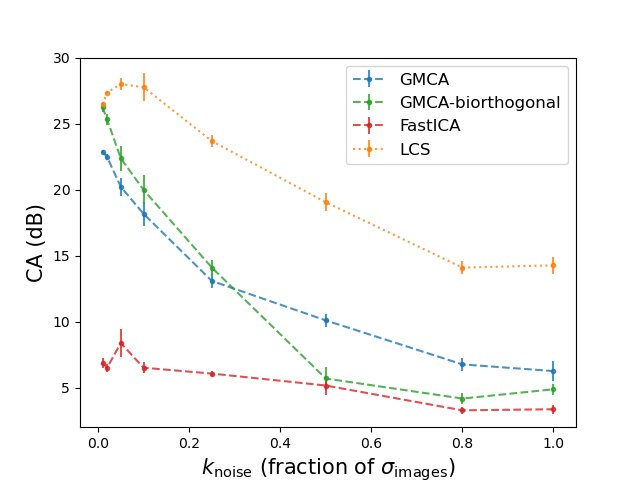}
    \caption{Same as Fig.~\ref{fig:fig3}, but for the multiple-source texture case: LCS is shown in orange, GMCA in blue, GMCA-biorthogonal in green, and FastICA in red.}
    \label{fig:fig9}
\end{figure*}

The resulting PSNR and CA are displayed in Fig.\ref{fig:fig9}, confirming the same trend observed in Section~\ref{sec:toy}: LCS consistently outperforms the other methods. The CA values are significantly higher across all noise levels, indicating improved source separation. However, this improvement does not fully translate to better $\mathbf{S}$ estimates here, likely due to the denoising capability of the Learnlet network’s final step, which plateaus around 20 dB at high noise levels, as illustrated in Fig.\ref{fig:fig2} showing the denoising performance.

In summary, optimizing the sparsity basis and threshold parameters individually for each source type—considering their specific morphologies, shapes, and features—further enhances the separation performance compared to using a common, global basis $\mathcal{L}$.

\begin{figure*}[!ht]
\centering
    \includegraphics[width=\linewidth]{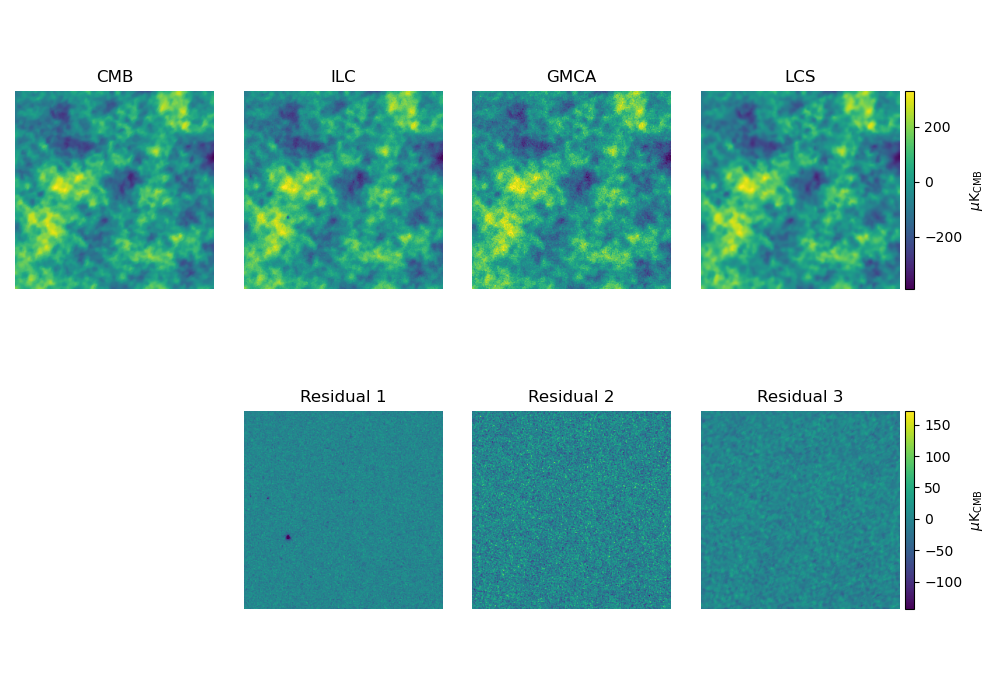}
    \caption{Results of different methods applied to a patch of the CMB. Top panel: reconstructed CMB patches. Bottom panel: residuals between the methods and the ground truth.}
    \label{fig:fig10}
\end{figure*}

\begin{figure*}[!ht]
\centering
    \includegraphics[width=\linewidth]{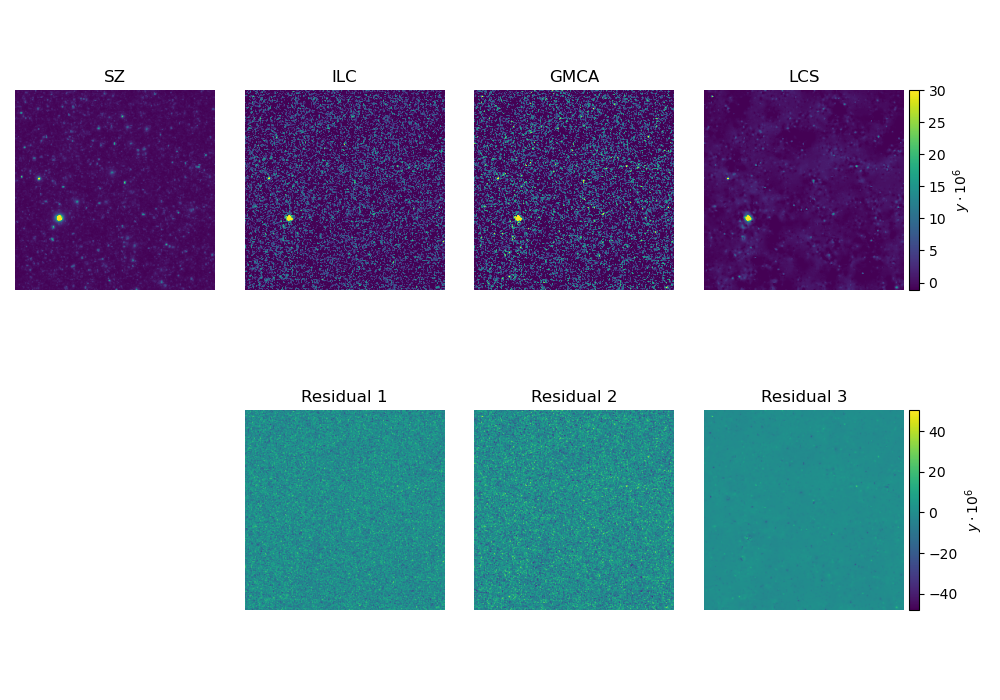}
    \caption{Same as Fig.~\ref{fig:fig10}, but for the patch of SZ.}
    \label{fig:fig11}
\end{figure*}

\subsubsection{Experiments with CMB and SZ data}

In this fourth experiment, we apply LCS to extract the CMB and SZ components from CMB data. Here, the weights for the CMB and SZ are theoretically known: the CMB weights are set to one across all frequencies in units of $\mu\mathrm{K}_\mathrm{CMB}$, while the SZ weights follow the frequency-dependent function in same units:
\begin{equation}
    f(x) = x \cdot \frac{e^x + 1}{e^x - 1} - 4,
\end{equation}
where $x = \frac{h \nu}{k_\mathrm{B} T_{\mathrm{CMB}}}$. Specifically, we chose the weights pre-computed from Websky Simulation presented in Table 2 in \cite{stein2020}.

The challenge in this case arises from the presence of numerous other components with unknown weights that can contaminate the signal, the most significant being the Cosmic Infrared Background (CIB) \citep{dole2006}, which is known to correlate with the SZ effect and contaminate SZ maps \citep{hurier2013, planck2016}. We then chose to include only the CMB, SZ, and CIB components in our experiments, using numerical simulations from WebSky \citep{stein2020} available online\footnote{\url{https://mocks.cita.utoronto.ca/data/websky/v0.0/}}. Frequency maps were constructed from five relevant Planck HFI channels: 100, 143, 217, 353, and 545 GHz. No beam effects or additional foreground contaminants were included, and white Gaussian noise was added with standard deviations estimated from the corresponding observed Planck frequency maps.

Before applying LCS, we trained individual Learnlet networks—similar to the approach described in the previous section—specifically designed to promote the sparsity of each expected component, denoted as $\mathcal{L}_\mathrm{CMB}$, $\mathcal{L}_\mathrm{SZ}$, and $\mathcal{L}_\mathrm{CIB}$. To construct the training data, we divided the full-sky HEALPix maps of each component (at $N_\mathrm{side}=2048$) into 3,092 overlapping square patches of $256\times256$ pixels, corresponding to an angular size of $\sim 7.2^\circ \times 7.2^\circ$ per patch. These patches cover the entire sky. We identified the patch with the highest SZ signal as the “target patch” for evaluation by selecting the patch with the highest SZ amplitude. This patch was excluded from the training process to avoid any overlap between training and testing data and the patch containing the highest SZ source signal is shown in Fig.~\ref{fig:fig11}. The Learnlet networks were then trained on the remaining 3,091 patches (hereafter referred to as the training set) for each component separately. The networks were trained in a supervised manner using the known clean component maps as ground truth. After training, we applied the LCS algorithm - using the trained $\mathcal{L}_\mathrm{CMB}$, $\mathcal{L}_\mathrm{SZ}$, and $\mathcal{L}_\mathrm{CIB}$ transforms - to the noisy frequency maps, restricted to the excluded target patch. This patch was used exclusively for evaluation and comparison with other methods (GMCA, ILC, and constrained ILC). The corresponding original CMB and SZ maps are shown in the left panels of Fig.~\ref{fig:fig10} and Fig.~\ref{fig:fig11}.

Table~\ref{tab:tab1} reports the PSNR obtained for all methods; we highlight in bold the best PSNR among the different methods on this patch. The corresponding reconstructed maps are shown in Fig.\ref{fig:fig10} and Fig.\ref{fig:fig11}. From the PSNR values in the patch, for the CMB case, GMCA appears to underperform relative to both ILC and LCS, achieving a PSNR of 26.28 dB. LCS delivers the best results, outperforming ILC by approximately 2 dB (35.65 dB for LCS vs. 32.88 dB for ILC). Visually, the residual maps reveal that LCS effectively removes SZ contamination, visible as “holes” in the recovered CMB maps. Regarding SZ, from PSNR values in the patch, ILC and GMCA yield comparable PSNR values of 27.90 dB and 25.56 dB respectively, with ILC performing slightly better. In contrast, LCS achieves a substantially higher PSNR of 42.94 dB, primarily due to effective noise suppression by the Learnlet denoiser $\mathcal{L}_\mathrm{SZ}$ applied as a post-processing step of LCS, as illustrated in Fig.~\ref{fig:fig11}. Overall, visually and from the PSNR, both the CMB and SZ maps are very well reconstructed using LCS.

However, unlike ILC—which is constructed to ensure that the residual map is (ideally) uncorrelated with the true signal—LCS does not impose such a constraint. As a consequence, some correlation between the residual and the true tSZ signal can remain, as visible in Fig.~\ref{fig:fig11} (bottom right). Although LCS provides a visually and quantitatively accurate reconstruction on this patch, this implies that additional work would be required to guarantee unbiased residuals for applications relying on external cross-correlations. This is a natural consequence of the design of LCS, which relies on a sparsity-driven prior instead of enforcing the unbiasedness constraint built into ILC. For cross-correlation analyses, where preserving all of the signal in the cleaned map is essential, an explicit debiasing step or constraint would need to be incorporated; we leave this for future work.

\begin{table}
\caption{\label{tab:tab1} Results of component separation in CMB data, showing the PSNR values for CMB and SZ reconstruction using ILC, GMCA, and LCS. The highest PSNR for each component is highlighted in bold to indicate the best reconstruction.}
\centering
\begin{tabular}{lcc}
\hline\hline
Method & CMB & SZ\\
\hline
ILC           & 32.88 & 27.90\\
GMCA           & 26.28   & 25.56\\
LCS & \textbf{34.65} & \textbf{42.94}\\
\hline
\end{tabular}
\end{table}

\section{Discussion and conclusion}\label{sec:sec4}

In this study, we introduced the Learnlet Component Separator (LCS), a new blind source separation framework that effectively integrates traditional sparsity-enforcing methods with the adaptability and expressivity of deep learning architectures. Utilizing the Learnlet transform—a learned, wavelet-inspired multiscale representation—LCS incorporates a data-driven prior within an iterative source separation procedure, delivering enhanced performance in complex astrophysical applications.

Our toy model numerical experiments show, quantified by PSNR values of extracted components compare to ground truth, that LCS generally outperforms traditional BSS methods like ICA, GMCA, and ILC, especially when dealing with non-Gaussian sources, complex spatial patterns, or varying noise levels, and especially at large noise - showing the robustness of learnlets to noise. The data-adaptive Learnlet transform allows the model to effectively capture the inherent structure of the sources, while its structured design provides both interpretability and computational efficiency—key advantages for handling large-scale datasets anticipated from future projects such as SKA. Although this proof-of-concept study has been made by quantifying the PSNR value only, we plan in future work to study in more detail contaminations and angular power spectra of the extracted components (Gkogkou et al., in prep.).

Several promising directions remain for future work. First, expanding the LCS framework to address nonlinear or convolutive mixing models \citep{decgmca} could extend its applicability to more realistic physical scenarios. For instance, simultaneously performing deconvolution and component separation will be essential to account for beam effects and evolving structures in radio experiments like SKA or Meerkat, enabling accurate extraction of the HI signal \citep{gkogkou2025}. Similarly, this will be critical for achieving the highest possible resolution in CMB or SZ maps in upcoming missions such as LiteBird or the Simons Observatory. Second, incorporating uncertainty quantification techniques could provide valuable measures of confidence in the separated components, which is vital for rigorous scientific interpretation. Finally, embedding domain-specific physical models or incorporating known spectral priors within the learning process may further improve both the robustness and interpretability of the results.

Beyond astrophysics, the methods developed in LCS have wide applicability across various signal processing fields where blind source separation is essential. This includes areas like biomedical imaging, remote sensing, audio band unmixing, and any discipline that requires interpretable and adaptable source separation techniques.

In conclusion, LCS paves the way for employing learnlets in blind source separation, providing a flexible, interpretable, and high-performing approach for complex signal decomposition challenges. While its hybrid architecture is especially well-suited for astrophysical applications, the underlying concepts readily extend to a broad range of modern signal processing domains.

\begin{acknowledgements}

The authors thank the anonymous referee for useful comments that contributed to a better version of this paper. The authors thank all members of the TITAN team for useful and fruitful discussions.
This work was supported by the TITAN ERA Chair project (contract no. 101086741) within the Horizon Europe Framework Program of the European Commission, and the Agence Nationale de la Recherche (ANR22-CE31-0014-01 TOSCA).
The sky simulations used in this paper were developed by the WebSky Extragalactic CMB Mocks team, with the continuous support of the Canadian Institute for Theoretical Astrophysics (CITA), the Canadian Institute for Advanced Research (CIFAR), and the Natural Sciences and Engineering Council of Canada (NSERC), and were generated on the Niagara supercomputer at the SciNet HPC Consortium (cite https://arxiv.org/abs/1907.13600). SciNet is funded by: the Canada Foundation for Innovation under the auspices of Compute Canada; the Government of Ontario; Ontario Research Fund - Research Excellence; and the University of Toronto.
\end{acknowledgements}

\bibliography{bibli.bib}

\begin{appendix}
\section{Interpreting the learnlets}

\begin{figure}[!ht]
\centering
    \includegraphics[width=\linewidth]{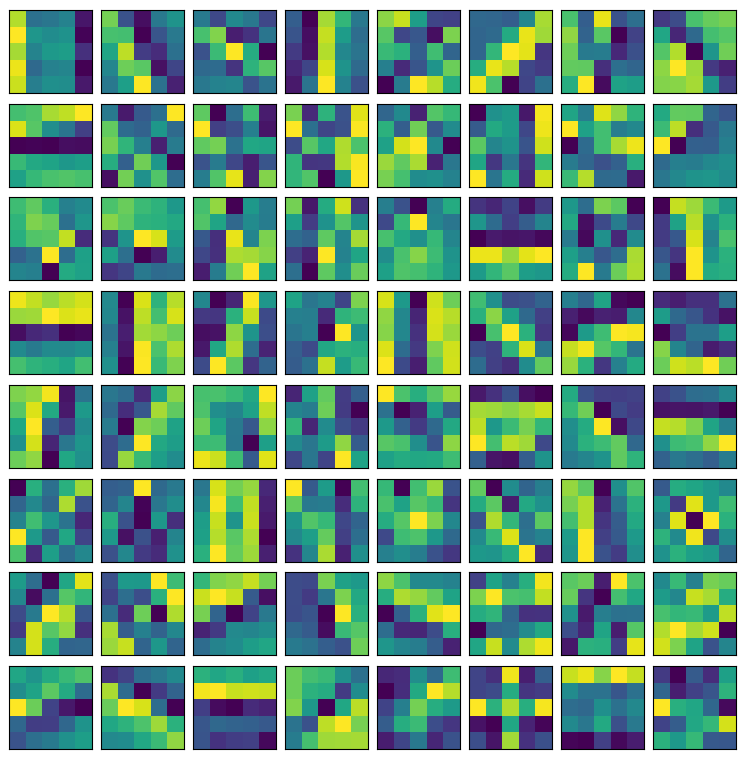}
    \caption{Example of learnlets from the first scale of the Learnlet network trained on ImageNet dataset in Sect.~\ref{sec:imagenet}. Distinct horizontal, vertical, and diagonal structures are clearly identifiable.}
    \label{fig:learnlets}
\end{figure}

The Learnlet network trained on the ImageNet dataset, as described in Sect.\ref{sec:imagenet}, exhibits interpretability. By examining the shapes of the learned filters—referred to as learnlets—we can identify key features and patterns that contribute to the sparse representation of the images. Figure\ref{fig:learnlets} displays the learnlets at the first scale, revealing various structures captured during training, such as horizontal, vertical, and diagonal orientations, along with cross-like and small circular patterns.

\begin{figure}[!ht]
\centering
    \includegraphics[width=\linewidth]{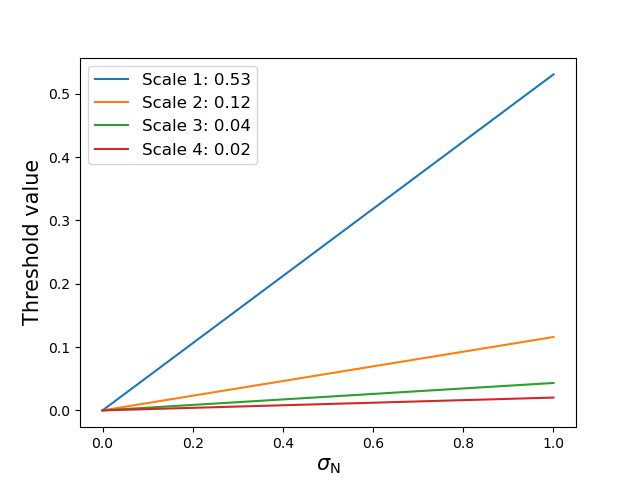}
    \caption{Threshold parameters $k_j$ learned during the training of the Learnlet network on the ImageNet dataset in Sect.~\ref{sec:imagenet}.}
    \label{fig:thresholds}
\end{figure}

The threshold parameters $k_j$ are also of particular interest. The values learned during training on the ImageNet dataset, as presented in Sect.\ref{sec:imagenet}, are shown in Fig.\ref{fig:thresholds}. Notably, these thresholds exhibit a decreasing trend, with higher values at finer scales and lower values at coarser scales. This pattern aligns with expectations, as Gaussian noise tends to be more prominent at fine scales and progressively diminishes at coarser levels, often becoming negligible at the coarsest scale. The network effectively captures this behavior during training through its learned thresholding mechanism.

\end{appendix}

\end{document}